\documentclass[12pt,a4paper,showkeys]{revtex4}
\usepackage{graphicx}
\usepackage{epsfig}
\usepackage{bm}
\usepackage{amsmath}
\usepackage{amssymb}
\usepackage{graphics}
\usepackage{epstopdf}
\usepackage[linkcolor=red,citecolor=green,urlcolor=blue]{hyperref}

\begin{document}

\title{ \textbf{Studies of backward particle production with A Fixed-Target
Experiment using the LHC beams}}
\author{Federico Alberto Ceccopieri}
\email{federico.alberto.ceccopieri@cern.ch}
\affiliation{IFPA, Universit\'e de Li\`ege,  All\'ee du 6 ao\^ut, \\
B4000, Li\`ege, Belgium}
\begin{abstract}
\noindent
The foreseen capability to cover the far backward region  
at A Fixed-Target Experiment using the LHC beams
allows to explore the dynamics of target fragmentation in hadronic collisions. 
In this report we briefly outline the required 
theoretical framework and discuss a number of studies of forward and backward particle production.
By comparing this knowledge with the one accumulated  
in Deep Inelastic Scattering on target fragmentation, the basic concept of QCD factorisation could
be investigated in detail. 
\end{abstract}

\maketitle

\section{Introduction}
\label{intro}
In hadronic collisions a portion of the produced particle spectrum 
is characterised by hadrons carrying a sizeable fraction of the available centre-of-mass energy,
the so-called leading particle effect. It is phenomenologically observed that for such hadrons their valence-parton composition is almost or totally conserved with respect to the one of inital-state hadrons~\cite{Basile}. 
In $pp$ collisions, for example, protons, neutrons and lambdas show a significant leading particle effect.
Moreover, for such semi-inclusive processes, the production cross section peaks at very small transverse momenta with respect to the collision axis, a regime where perturbative techniques can not be applied, 
giving insight on non-perturbative aspects of QCD dynamics in high energy collisions.

Quite interestingly, the leading particle effect has been observed in Semi-Inclusive Deep Inelastic Scattering (SIDIS).
At variance with the hadronic processes mentioned above, such a process naturally involves a large momentum transfer. 
The presence of a hard scale enables the derivation of a dedicated factorisation theorem~\cite{factorisation_soft,factorisation_coll} which ensures  that QCD factorisation holds for backward particle production in DIS. 
The relevant cross sections can then be factorised into perturbatively calculable short-distance cross sections and new distributions, fracture functions, which simultaneously encode information both on the
interacting parton and on the spectator fragmentation into the observed hadron. 
Despite of being non-perturbative in nature, their scale dependence can be calculated within perturbative QCD~\cite{trentadue_veneziano}.
The factorisation theorem~\cite{factorisation_soft,factorisation_coll} 
guarantees that fracture functions are universal distributions, at least in the context of SIDIS.

Detailed experimental studies of hard diffraction at HERA have shown to support 
the hypothesis of QCD factorisation and evolution inherent the fracture functions formalism. Furthermore 
they led to a quite accurate knowledge of diffractive parton distributions~\cite{H106LRG,H107dijet,ZEUS09final,myDPDF},
a special case of fracture functions in the very backward kinematic limit. 
For particles other than protons, proton-to-neutron fracture functions have been extracted from 
a pQCD analysis of forward neutron production in DIS in Ref.~\cite{ceccopieri_neutron}. 
A set of proton-to-lambda fracture functions has been obtained by performing a combined pQCD fit to a variety of 
semi-inclusive DIS lambda production data in Ref.~\cite{ceccopieri_mancusi}. 

As theoretically anticipated in Ref.~\cite{break1,break2,factorisation_soft}
and experimentally observed in hard diffraction in $p \bar{p}$ collisions at Tevatron~\cite{break_exp1,break_exp2},
QCD factorisation is violated for fracture functions in hadronic collisions. 
On general grounds, it might be expected, in fact, that the dynamics of target-remnants hadronisation is affected by 
the coloured environment resulting from the scattering in a rather different way with respect to 
the Deep Inelastic Scattering case. 

Nonetheless, the tools mentioned above allow to investigate 
quantitatively particle production mechanisms in the very backward and 
forward regions, to test the concept of factorisation at the heart of QCD 
and to study the dependencies of factorisation breaking 
upon the species and the kinematics of the selected final state particle.

This physics program could be successfully carried on at 
A Fixed-Target Experiment using the LHC beams~\cite{AFTER}. 
Novel experimental techniques are, in fact, available to extract
beam-halo protons or heavy-ions from LHC beams without affecting LHC performances.
Such a resulting beam would be then impinged on a high-density and/or long-lenght
fixed target, guaranteeing high luminosities. 
Furthermore and most importantly for the physics program to be discussed here, 
the entire backward hemisphere (in the centre-of-mass system of the collision)
would be accessible with standard experimental techniques allowing high precision studies of target fragmentation.
Althought measurements of particle production in the very forward region (close to the beam axis) 
might be challenging experimentally due to the high particle densities and large energy flow, 
the installation of dedicated detectors, like forward neutron calorimeters and/or 
proton taggers, could further broaden the physics program oulined above
giving access to the beam fragmentation region.

The paper is organised as follows. In Sec.~\ref{sec:1} we first give a brief theoretical introduction 
on the fracture functions formalism and to higher order corrections to the 
semi-inclusive Drell-Yan process. In Sec.~\ref{sec:2} we outline different analyses which could
performed at AFTER@LHC with special focus on single hard diffraction. In Sec.~\ref{Conclusions} we summarise our results.

\section{Collinear factorisation formula}
\label{sec:1}
Fracture functions, originally introduced in DIS, do depend on a large momentum transfer. 
Therefore, in order to use them in hadronic collisions, a hard process must be selected.
We consider here the semi-inclusive version of the Drell-Yan process, 
\begin{equation}
H_1(P_1) + H_2(P_2) \rightarrow H(h) + \gamma^*(q) + X, 
\end{equation}
in which one hadron $H$ is measured in the final 
state together with a Drell-Yan pair. 
In such a process the high invariant mass of the lepton pair, $q^2=Q^2$, allows the applicability of perturbative QCD,
while the detected hadron $H$ can be used, without any phase space restriction,
as a local probe to investigate particle production mechanisms.

\begin{figure}[t]
\centerline{\includegraphics[scale=1.0]{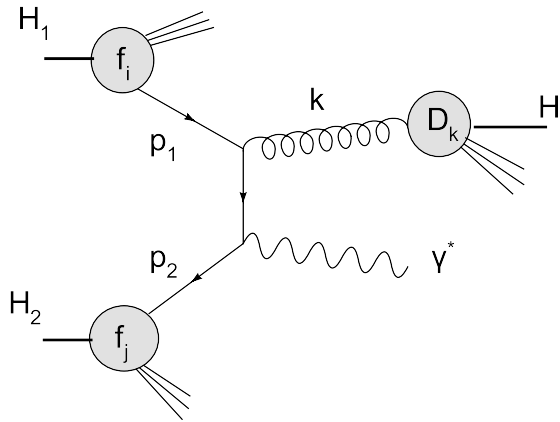}}
\caption{\textsl{Example of diagram contributing to hadron production in the central
fragmentation region to order $\mathcal{O}(\alpha_s)$ in eq.(\ref{NLOcentral}).}}
\label{fig1}
\end{figure}

The associated production of a particle and a Drell-Yan pair 
in term of partonic degrees of freedom starts 
at $\mathcal{O}(\alpha_s)$. One of the contributing diagrams 
is depicted in Fig.~(\ref{fig1}). 
Assuming that the hadronic cross-sections 
admit a factorisation in term of long distance non-perturbative 
distributions and short distance perturbative calculable matrix elements 
for the partonic process $i(p_1)+j(p_2)\rightarrow l(k)+\gamma^*(q)$, 
predictions based on perturbative QCD are obtained 
convoluting the relevant partonic sub-process cross-sections, $d\hat{\sigma}^{ij \rightarrow l \gamma^* }$, 
with parton distribution functions, $f_{i/H_1}$ and $f_{j/H_2}$, and fragmentation functions,
$D^{H/l}$. The hadronic cross section, 
at centre of mass energy squared $s=(P_1+P_2)^2$, 
can be symbolically written as~\cite{SIDYmy,SIDYmy2}
\begin{equation}
\label{NLOcentral}
\frac{d\sigma^{H,C,(1)}}{dQ^2 dz} \propto \; \sum_{i,j,l}
\int \frac{dx_1}{x_1} \int \frac{dx_2}{x_2} \int \frac{d\rho}{\rho}
 f_i^{[1]}(x_1) \, f_{j}^{[2]}(x_2)  \, D^{H/l}(z/\rho) \,
\frac{d\hat{\sigma}^{ij\rightarrow l\gamma^*}}{dQ^2 d\rho},
\end{equation}
where the convolutions are over the momentum fractions  
of the incoming and outgoing partons. The partonic indeces $i$, $j$ and $l$ in the sum 
run on the available partonic subprocesses. 
The superscripts label the incoming hadrons and the presence of crossed terms is understood.
This type of factorised hadronic cross section is expected to hold
for hadrons produced at sufficiently high transverse momentum and it is widely and successfully used to 
compute cross sections for large momentum transfer processes in hadronic collisions.
The Lorentz-invariant variable $z$ in eq.~(\ref{NLOcentral}) is defined by
\begin{equation}
\label{z}
z=\frac{2h\cdot(P_1+P_2)}{s}\equiv \frac{2E_H^*}{\sqrt{s}}\,.
\end{equation}
In the hadronic centre-of-mass frame, where the second identity holds, 
$z$ is just the observed hadron energy, $E_H^*$, scaled down by the beam energy $\sqrt{s}/2$. 
The variable $\rho$, appearing in eq.~(\ref{NLOcentral}), is its partonic equivalent. 
Within this production mechanism, the observed hadron $H$ is 
generated by the fragmentation  of the final state parton $l$, and for this reason 
we address it as \textit{central}. The amplitudes squared~\cite{DYNLO}, however,
are singular when the transverse momentum of the final state parton vanishes. 
In such configurations, the parent parton $l$ of the
observed hadron $H$ is collinear either to the incoming parton $i$ or $j$.
As these phase space region are approached, perturbation theory looses its predictivity. 
This class of collinear singularities 
escape the usual renormalisation procedure which amounts to reabsorb  collinear divergences 
into a redefinition of bare parton distribution and fragmentation functions.
Such singularities are likely to appear in every fixed order calculation
in the same kinematical limits spoiling the convergence of the perturbative 
series. In Refs~\cite{SIDYmy,SIDYmy2} a generalised procedure for the factorisation of such additional collinear singularities is proposed. The latter is the same as the one proposed in Deep Inelastic Scattering~\cite{graudenz}
where the same singularities pattern is also found, 
confirming the universality of collinear radiation between different hard processes.  
Such a generalised collinear factorisation makes use of fracture functions. 
These distributions obey DGLAP-type evolution equations which contain an additional inhomogeneous term 
resulting from the subtraction of collinear singularities in the target-fragmentation region~\cite{trentadue_veneziano,graudenz}. Such equations allow to resum the corresponding large logarithmic corrections to all orders in perturbation theory. Bare fracture functions, $M^{H/H_1}_i(x,z)$, describe 
the hadronization of the spectators system in hadron-induced reactions.
They express the conditional probability 
to find a parton $i$ entering the hard scattering while an hadron $H$ 
is produced with fractional momentum $z$ in the target fragmentation region 
of the incoming hadron $H_1$. 

The use of fracture functions allows  
particle production already to $\mathcal{O}(\alpha_s^0)$, 
since the hadron $H$ can be non-pertubatively produced 
by a fracture function itself.
\begin{figure}[t]
\centerline{\includegraphics[scale=1.0]{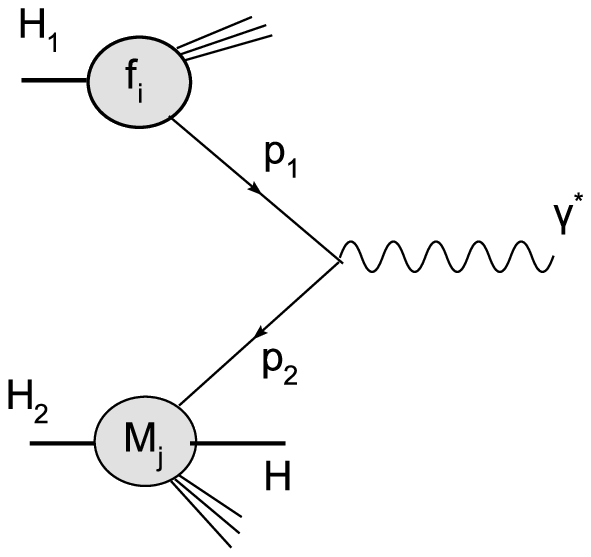}
\includegraphics[scale=1.0]{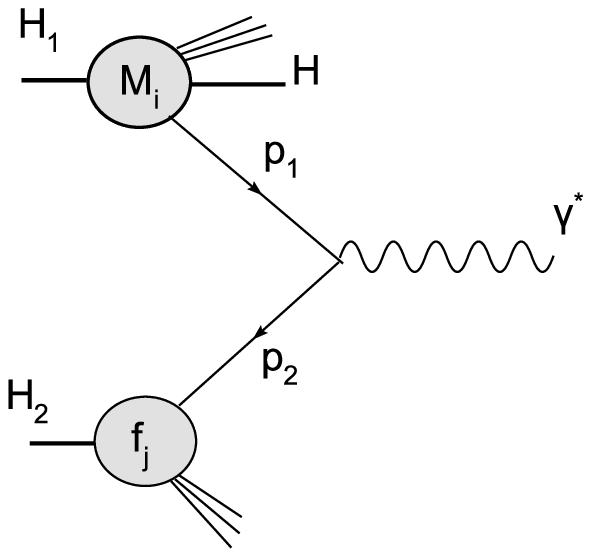}}
\caption{\textsl{Pictorial representation of the parton model formula, eq.~(\ref{LO}), for the associated production of a Drell-Yan pair and a particle in the target fragmentation regions.}}
\label{fig2}
\end{figure}
Therefore the lowest order parton model formula can be symbolically written as
\begin{equation}
\label{LO}
\frac{d\sigma^{H,T,(0)}}{dQ^2 dz} \propto \sum_{i,j}
\int \frac{dx_1}{x_1} \int \frac{dx_2}{x_2} 
\big[M_i^{[1]}(x_1,z) \, f_j^{[2]}(x_2) + M_i^{[2]}(x_2,z) \, f_j^{[1]}(x_1)\big] 
\frac{d\hat{\sigma}^{ij\rightarrow \gamma^*}}{dQ^2}
\end{equation} 
and it is sketched in Fig.~(\ref{fig2}).
The superscripts in eq.~(\ref{LO}) indicate from which incoming hadron, $H_1$ or $H_2$,
the outgoing hadron $H$ is produced through a fracture functions.
In order to complete the calculation to $\mathcal{O}(\alpha_s)$ accuracy
we should consider higher order corrections to eq.~(\ref{LO}). 
Since in this case the hadron $H$ is already produced by fracture functions,
final state parton radiation should be 
integrated over and the resulting contribution added to virtual corrections.
One of the contributing diagrams is depicted in Fig.~(\ref{fig3}).
\begin{figure}[t]
\centerline{\includegraphics[scale=1.0]{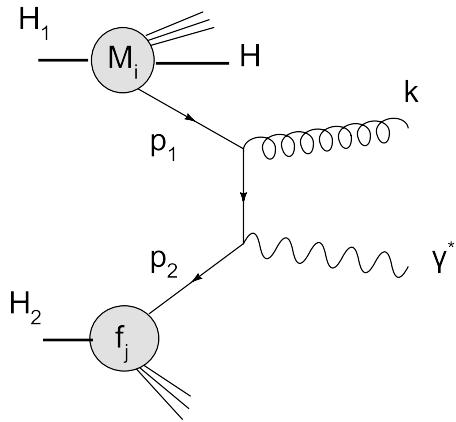}}
\caption{\textsl{Example of diagram contributing to $\mathcal{O}(\alpha_s)$ corrections in the 
target fragmentation region, eq.(\ref{NLOtarget}).}}
\label{fig3}
\end{figure}
The general structure of these terms is  
\begin{equation}
\label{NLOtarget}
\frac{d\sigma^{H,T,(1)}}{dQ^2 dz} \propto \sum_{i,j}
\int \frac{dx_1}{x_1} \int \frac{dx_2}{x_2} 
\big[M_i^{[1]}(x_1,z) \, f_j^{[2]}(x_2) + M_i^{[2]}(x_2,z) \, f_j^{[1]}(x_1)\big] 
\frac{d\hat{\sigma}^{ij\rightarrow (l)\gamma^*}}{dQ^2}\,.
\end{equation}
We refer to them as to the \textit{target} fragmentation contributions.
Their calculation is, a part from minor differences in kinematics, completely  
analogous to the one of the inclusive Drell-Yan case. The factorisation procedure, first elaborated in Ref.~\cite{graudenz} in the context of SIDIS, amounts to substitute in eq.~(\ref{LO}) the bare
fracture and parton distributions functions with their renormalised version~\cite{SIDYmy,SIDYmy2}. 
Renormalised parton distributions and fracture functions 
homogeneous terms do cancel, as in the inclusive Drell-Yan case, all singularities present in eq.~(\ref{NLOtarget}).
The additional singularities in eq.~(\ref{NLOcentral}) are cancelled by the combination 
of parton distributions and fracture functions inhomogeneous renormalisation terms.
Adding all the various contributions, the resulting hadron-$p_t$ integrated 
cross section, up to order $\mathcal{O}(\alpha_s)$, is then infrared finite~\cite{SIDYmy,SIDYmy2} 
and can be simbolically written as   
\begin{multline}
\label{NLOtot}
\frac{d\sigma^H}{dQ^2 dz} \propto \frac{\sigma_0}{N_c s} \sum_{i,j}
\big[M_i^{[1]} \otimes f_j^{[2]} + M_i^{[2]} \otimes f_j^{[1]} \big]
\big(1+\frac{\alpha_s}{2\pi} C^{ij}\big)+ \\
+\frac{\sigma_0}{N_c s} \; \frac{\alpha_s}{2\pi} \; \sum_{i,j,l} f_i^{[1]} \otimes f_{j}^{[2]} \otimes D^{H/l} 
\otimes K_{l}^{ij},
\end{multline}  
where $\sigma_0=4\pi\alpha_{em}^2/3 Q^2$ and $N_c$ is the number of colors.
We refer to the previous equation as to the collinear factorisation formula 
for the process under study.
The next-to-leading order coefficients $C^{ij}$ and $K_{l}^{ij}$ have been calculated in Ref.~\cite{SIDYmy2}, 
making the whole calculation ready for numerical implementation.  

We stress, however, that our ability to consistently subtract collinear singularities in such a semi-inclusive 
process is a necessary but not sufficient condition for factorisation to hold in hadronic
collisions. The one-loop calculation outlined above in fact does involve only the so-called active partons. It completely ignores multiple soft parton exchanges between active and spectators partons, 
whose effects should be accounted for in any proof of QCD factorisation. 
Therefore there is no guarantee that fracture functions extracted from SIDIS can be 
successfully used to describe forward or backward particle production in hadronic collisions.   
Reversing the argument, such a comparison may instead offer new insights on non-perturbative
aspects of QCD and to the breaking of factorisation. 

\section{Single hard diffraction at AFTER@LHC}
\label{sec:2}
As an application of the formalism presented in the previous sections
we will consider single hard diffractive production of a Drell-Yan pair
\begin{equation}
\label{single_hard_diff}
p_1 (P_1) + p_2 (P_2) \rightarrow p(P) + \gamma^*(q) +X\,,
\end{equation} 
where we have indicated in parathesis the four momenta of the relevant particles.
We present in the following cross sections differential in the virtual photon variables.
The subsequent decay of the virtual photon into a lepton pair can be easily 
included so that realistic cuts on leptons rapidity and transverse momentum can then be applied. 
We consider the AFTER@LHC kinematic setting in which a 7 TeV proton 
beam collides on a fixed target proton leading to a centre-of-mass energy 
of $\sqrt{s}=115$ GeV. We consider the projectile proton $p_1$ moving in the positive 
$z$ direction and $p_2$ at rest in the laboratory. The diffractively produced proton $p$ 
has in general almost the incoming projectile proton $p_1$ energy and very small 
transverse momentum as measured with respect to the collision axis. The detection 
of such fast protons will in general require the installation of forward proton taggers.
The lepton pair instead will be measured by the main AFTER@LHC detector. 
This kinematical configurations is pictorially represented in right plot of Fig.~(\ref{fig2}).

Diffractive processes has been intensively analysed in DIS at HERA $ep$ collider, 
revealing their leading twist nature. From scaling violations of the diffractive 
structure functions~\cite{H106LRG,myDPDF} and dijet production in the final state~\cite{H107dijet,ZEUS09final} 
quite precise diffractive parton distributions functions (dPDFs) have been extracted from HERA data, which 
parametrise the parton content of the color singlet exchanged in the $t$-channel.
The comparison of QCD predictions for single diffractive hard processes
based on diffractive parton distributions measured at HERA  
against data measured at Tevatron~\cite{break_exp1,break_exp2} ($\sqrt{s}=1.96$ TeV), 
adopting a factorised ansatz as in eq.~(\ref{LO}), 
has indeed revealed that these processes are, not unexpectedly ~\cite{break1,break2},
significantly suppressed in hadronic collisions.
This conclusion persists even after the inclusion of higher order QCD corrections~\cite{klasen}.
Complementing these results with the forthcoming ones at LHC at higher at $\sqrt{s}=13$ TeV   
and the ones from AFTER@LHC at $\sqrt{s}=115$ GeV will give insight on the energy dependence of the socalled
rapidity gap survival (RGS) probability in a wide range in $\sqrt{s}$. Since the theoretical computation 
of the RGS factor is highly model dependent, we decided not to include it
in our predicted cross sections which must be considered therefore as upper bounds.  

Diffractive parton distributions $f_i^D$ are in general proton-to-proton 
fracture functions $M_i$. They depend upon the final state proton fractional energy 
loss, $x_{\!I\!P}=1-z$ with $z$ given in eq.~(\ref{z}), the fractional momentum 
of the interacting parton with respect to the pomeron momentum, $\beta=x/x_{\!I\!P}$ and the 
virtuality $Q^2$. In general fracture functions may depend also upon 
the invariant momentum transfer $t=(P-P_1)^2$ at the proton vertex~\cite{extendedM}. 
In all diffractive structure functions measuraments at HERA, out of which dPDFs are determined, 
$t$ is integrated over up to some $t^{max} \ll Q^2$. In this case dPDFs 
obey ordinary DGLAP evolution equations~\cite{newfracture} as their extended, $t$-dependent, version~\cite{extendedM}.
In the present paper we use dPDFs form Ref.~\cite{H106LRG} which are defined 
by $|t|<1$ Ge$\mbox{V}^2$. Since they are extracted from large rapidity gap data
where the proton is not directly measured, they contain a contribution (23\%) from the socalled proton dissociation 
contribution. In order to use dPDFs in the present context we first note that 
\begin{equation}
\label{M_to_fD}
M_i(x_1,z,Q^2)=x_{\!I\!P}^{-1} f_i^D(\beta,x_{\!I\!P},Q^2)\,.
\end{equation}
The extra factor $x_{\!I\!P}^{-1}$ comes from the jacobian of the 
change $x_1= \beta x_{\!I\!P}$. 
We then rearrange eq.~(\ref{LO}) in terms of new variables obtaining
\begin{equation}
\label{LO_diff_inty}
\frac{d\sigma^{D}}{dQ^2 dx_{\!I\!P}} = \frac{\sigma_0}{N_c s} 
\int_{\tau/\beta}^{1} \frac{d\beta}{\beta} \sum_{q,\bar{q}} e_q^2
x_{\!I\!P}^{-1} f_q^D(\beta,x_{\!I\!P},\mu_F^2)
\, f_{\bar{q}}\Big(\frac{\tau}{\beta x_{\!I\!P}},\mu_F^2\Big) \,,
\end{equation} 
with $\tau=Q^2/s$. For simplicity we consider here leading order formulas but the extension to higher 
order is straightforward. In eq.~(\ref{LO_diff_inty}) we use parton distribution functions 
from Ref.~\cite{mrst01}. We show explicitely the dependence of fracture and parton distributions functions upon the factorisation scale, $\mu_F^2$. Predictions are obtained with this scale set to $\mu_F^2=Q^2$. Theoretical errors associated with higher order corrections are instead estimated 
varying such scale in the range $\mu_F^2=1/4 Q^2$ and $\mu_F^2=4 Q^2$. 

\begin{figure}[t]
\includegraphics[scale=0.6]{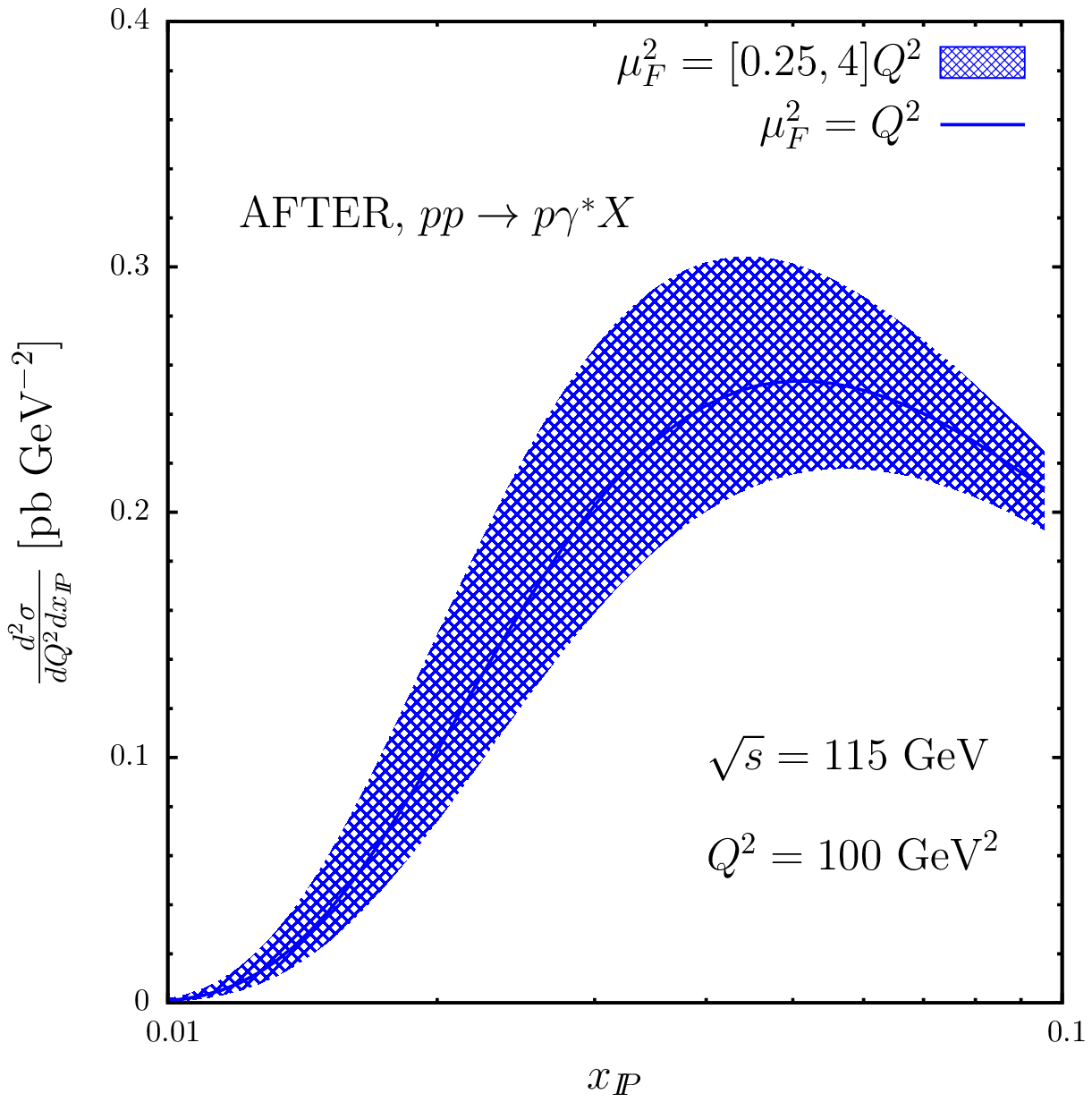}
\includegraphics[scale=0.6]{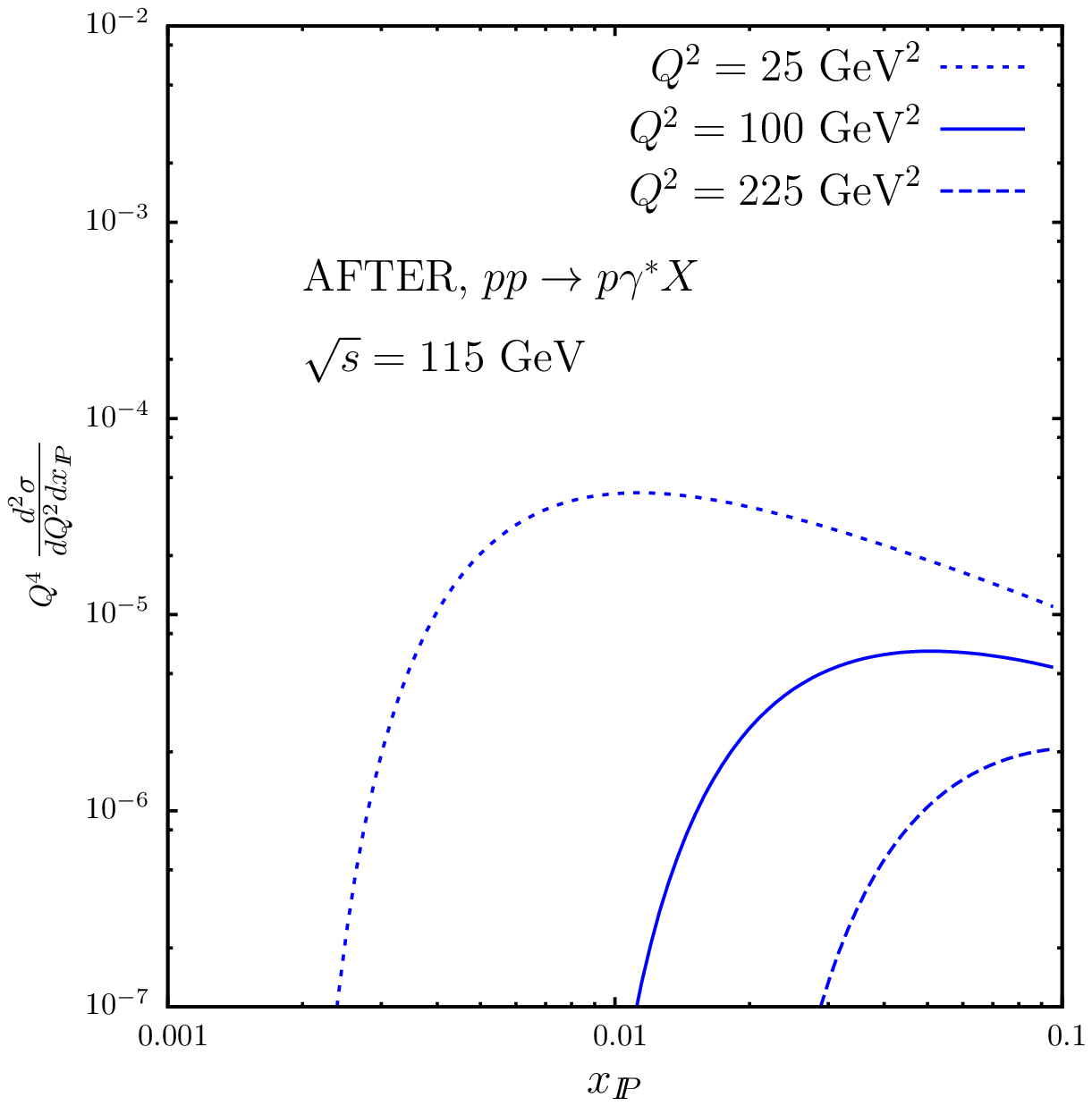}
\caption{\textsl{Left: Double differential cross sections for the production of a Drell-Yan pair of mass $Q^2=100$ Ge$\mbox{V}^2$. Blue error bands represent theoretical errors estimation, as described in the text. Right: Double differential cross sections times $Q^4$ for three different invariant masses.}}
\label{fig:xpom}
\end{figure}

In Fig.~(\ref{fig:xpom}) we present predictions for the $x_{\!I\!P}$ distribution.
In left plot we consider a Drell-Yan pair of mass $Q^2=100$ Ge$\mbox{V}^2$. 
The distribution shrinks as lower $x_{\!I\!P}$-values are approached whereas, 
from hard diffraction at HERA, it is well known that diffractive cross sections rise as an inverse power of $x_{\!I\!P}$. Such effect therefore is then attributed to phase space threshold effects. 
The Drell-Yan invariant mass constraint can be rewritten in the diffractive case as $Q^2 = \beta x_{\!I\!P} x_2 s$, which can be cast (for $\beta \rightarrow 1$ and $x_2\rightarrow 1$) in upper bound on the invariant mass $Q^2 < x_{\!I\!P} s $ at fixed  $x_{\!I\!P}$ and $s$. This hypothesis is further supported in the 
right plot of Fig.~(\ref{fig:xpom}), where differential distributions are presented for three 
values of $Q^2$. The former is multiplied by $Q^4$ to compensate the fast fall off of the electromagnetic cross section.
The lowest values of $x_{\!I\!P}$ are then accessed only by lowering the invariant mass of the pair. 
We note that, even considering the maximum value of $x_{\!I\!P}=0.1$, single diffractive 
production of $W^\pm$ and $Z$ is beyond the kinematic reach at AFTER@LHC. 
\begin{figure}[t]
\includegraphics[scale=0.6]{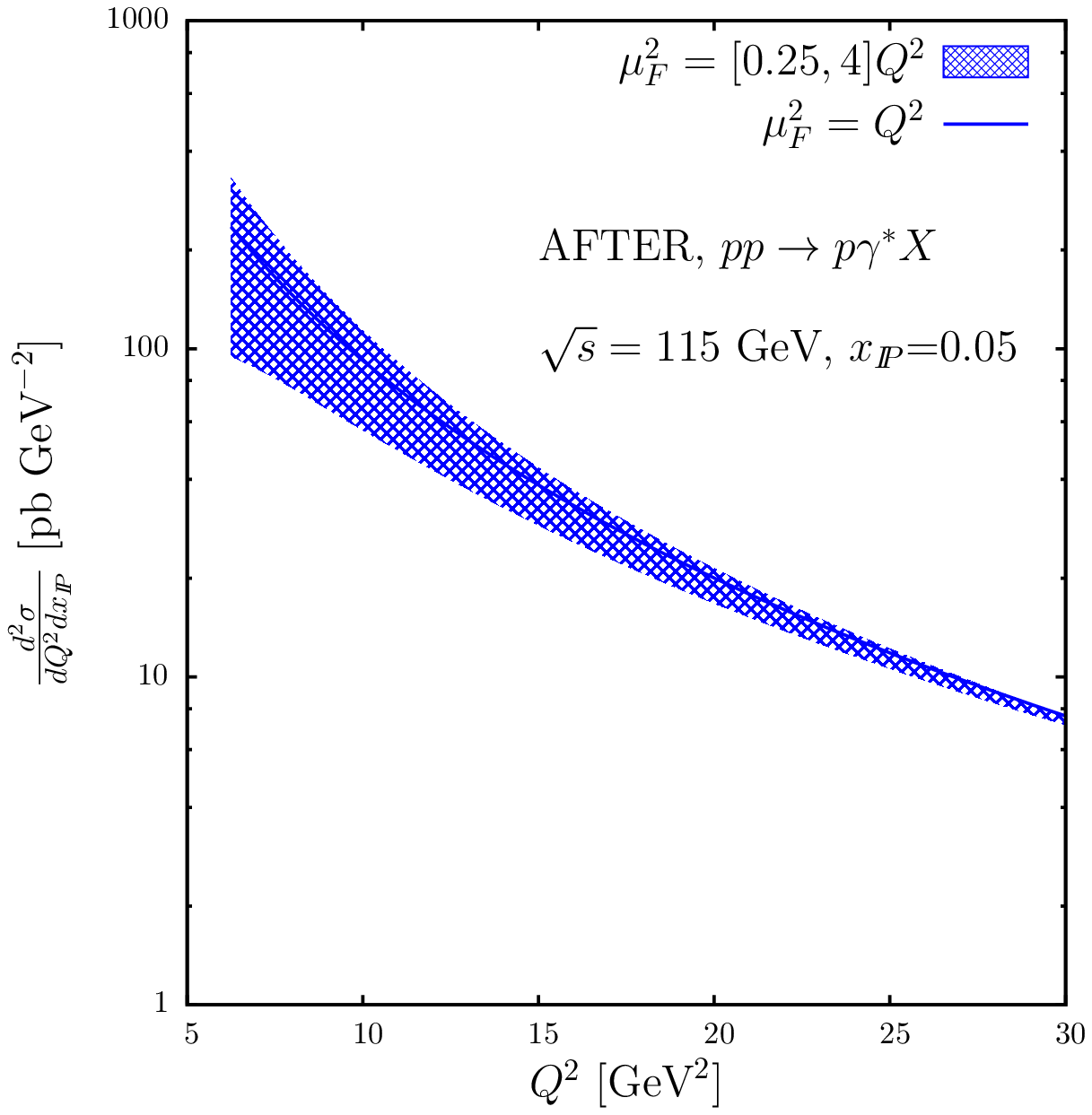}
\includegraphics[scale=0.6]{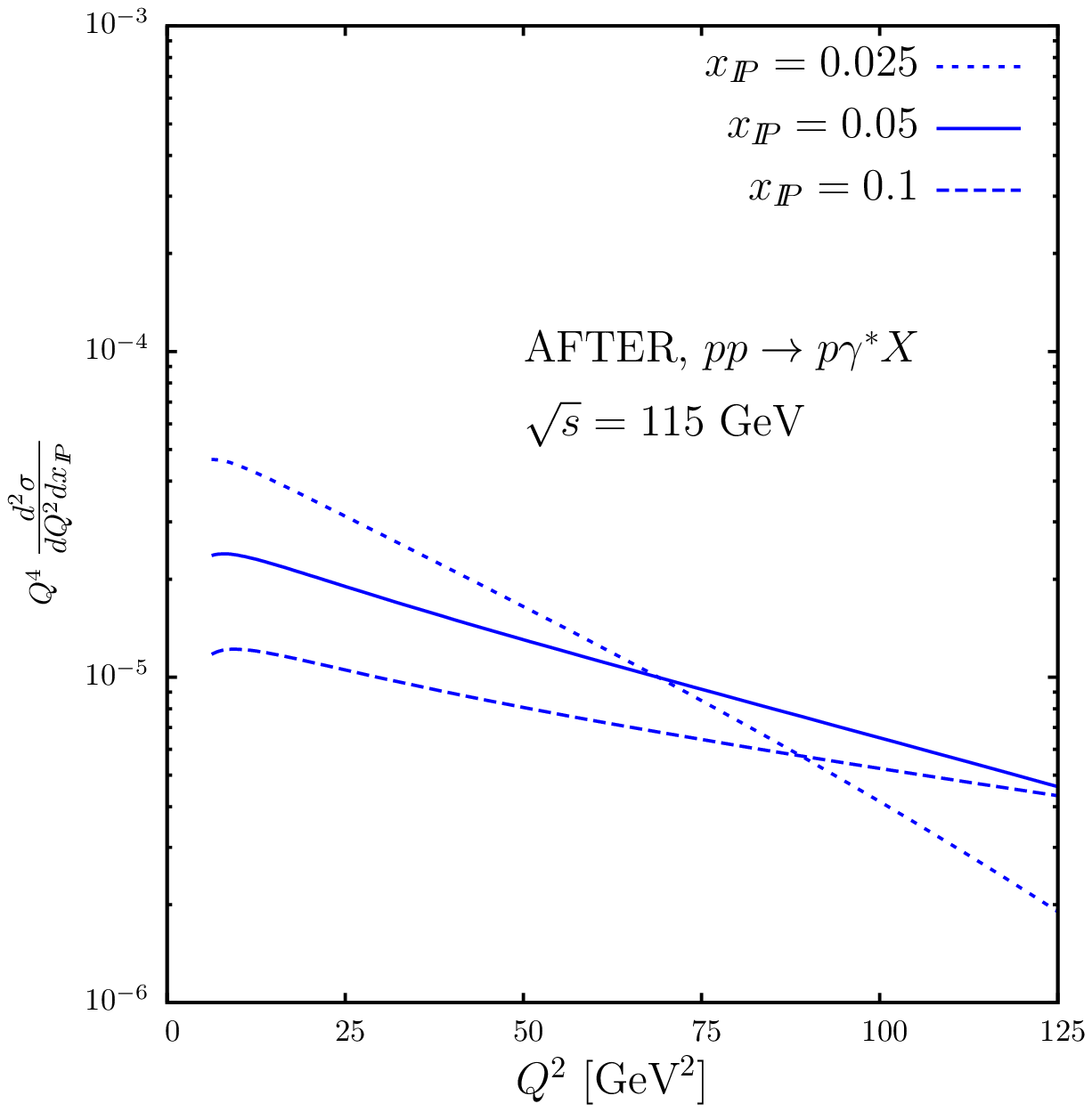}
\caption{\textsl{Left: Double differential cross sections for the production of a Drell-Yan pair at $x_{\!I\!P}=0.05$. Blue error bands represent theoretical errors estimation, as described in the text. Right: Double differential cross sections times $Q^4$ for three different $x_{\!I\!P}$-values.}}
\label{fig:Q2}
\end{figure}
In the left panel of Fig.~(\ref{fig:Q2}) we present the prediction for the $Q^2$ distribution at a fixed value 
of $x_{\!I\!P}=0.05$. The cross section, as expected, is fast falling as an inverse power of $Q^2$. 
The $Q^2$ distribution is particularly instructive since it allows to study
the possible dependence of the RGS factor on $Q^2$ and therefore to determine the underlying dynamics.  
In the right panel of Fig.~(\ref{fig:Q2}) we present the $Q^2$-differential cross section again 
multiplied by the factor $Q^4$. In this way all the $Q^2$ dependence is accounted for by 
that of fracture and parton distributions. These curves and the corresponding slopes, however, can not 
be readily interpreted as genuine results of QCD evolution of fracture and parton distributions functions
because of the threshold effect mentioned above appearing at such moderate values of $\sqrt{s}$.

By changing variable from $\beta$ to the virtual photon centre-of-mass rapidity, $y^{CM}$,  
\begin{equation}
\label{rapidity}
\beta=\frac{\sqrt{\tau}}{x_{\!I\!P}} \, e^{y^{CM}}, \;\; x_2= \sqrt{\tau} \, e^{-y^{CM}}\,,
\end{equation}
eq.~(\ref{LO_diff_inty}) can be further manipulated
to give the three-differential cross section 
\begin{equation}
\label{LO_diff_diffy}
\frac{d\sigma^{D}}{dQ^2 dx_{\!I\!P} dy^{CM}} = \frac{\sigma_0}{N_c s} 
\sum_{q,\bar{q}} e_q^2 x_{\!I\!P}^{-1} f_q^D(\beta,x_{\!I\!P},\mu_F^2)
\, f_{\bar{q}}(x_2,\mu_F^2) \,.
\end{equation} 
The rapidity range for diffractive Drell-Yan production reads
\begin{equation}
\ln \sqrt{\tau} < y^{CM} < \ln \frac{\sqrt{\tau}}{x_{\!I\!P}} 
\end{equation}
which, as expected, turns out to be asymmetric given the kinematic constraint $x_1<x_{\!I\!P}$. The rapidity range for the inclusive Drell-Yan case 
is recovered simply setting $x_{\!I\!P}=1$.
\begin{figure}[t]
\includegraphics[scale=0.6]{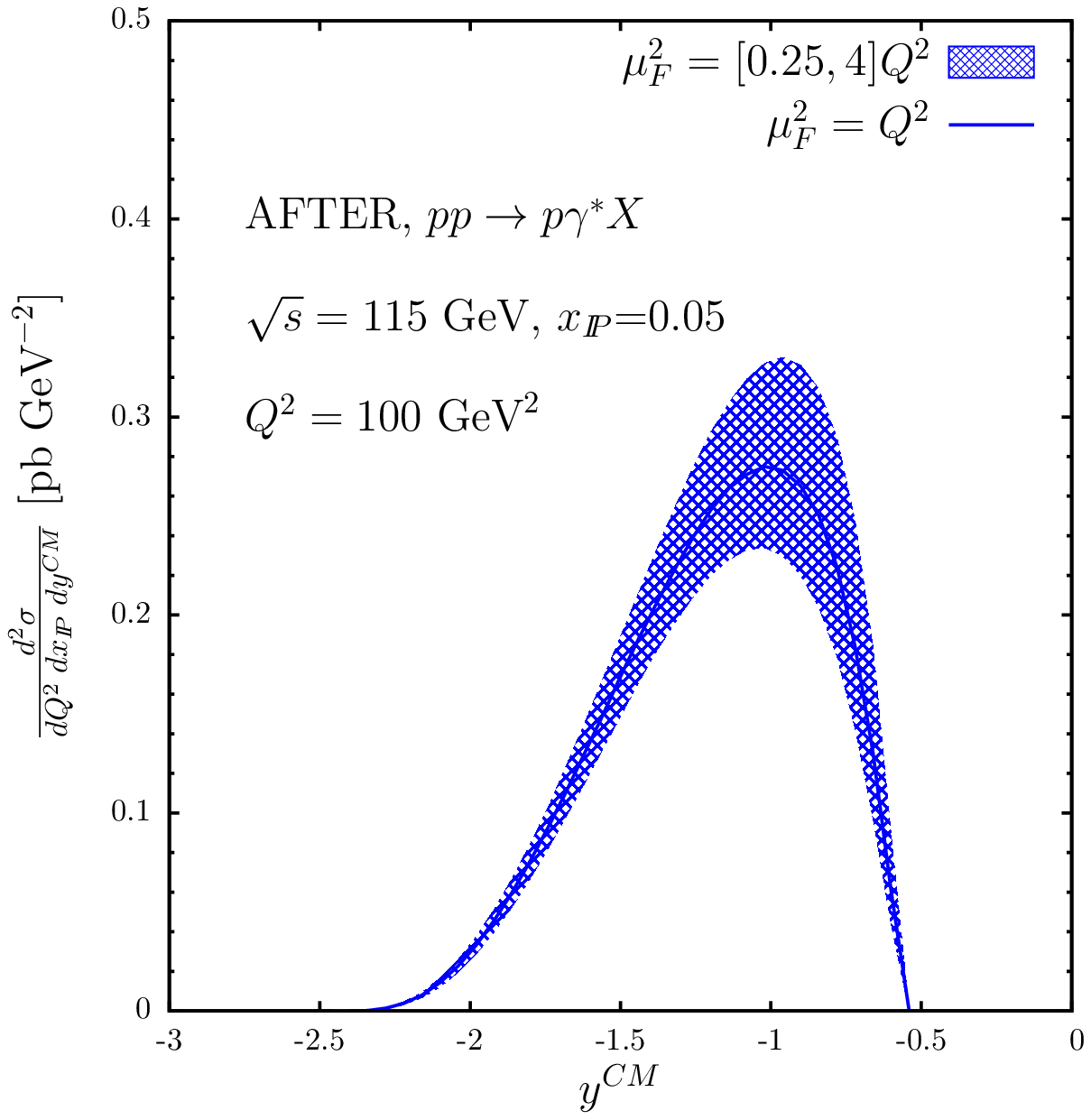}
\includegraphics[scale=0.6]{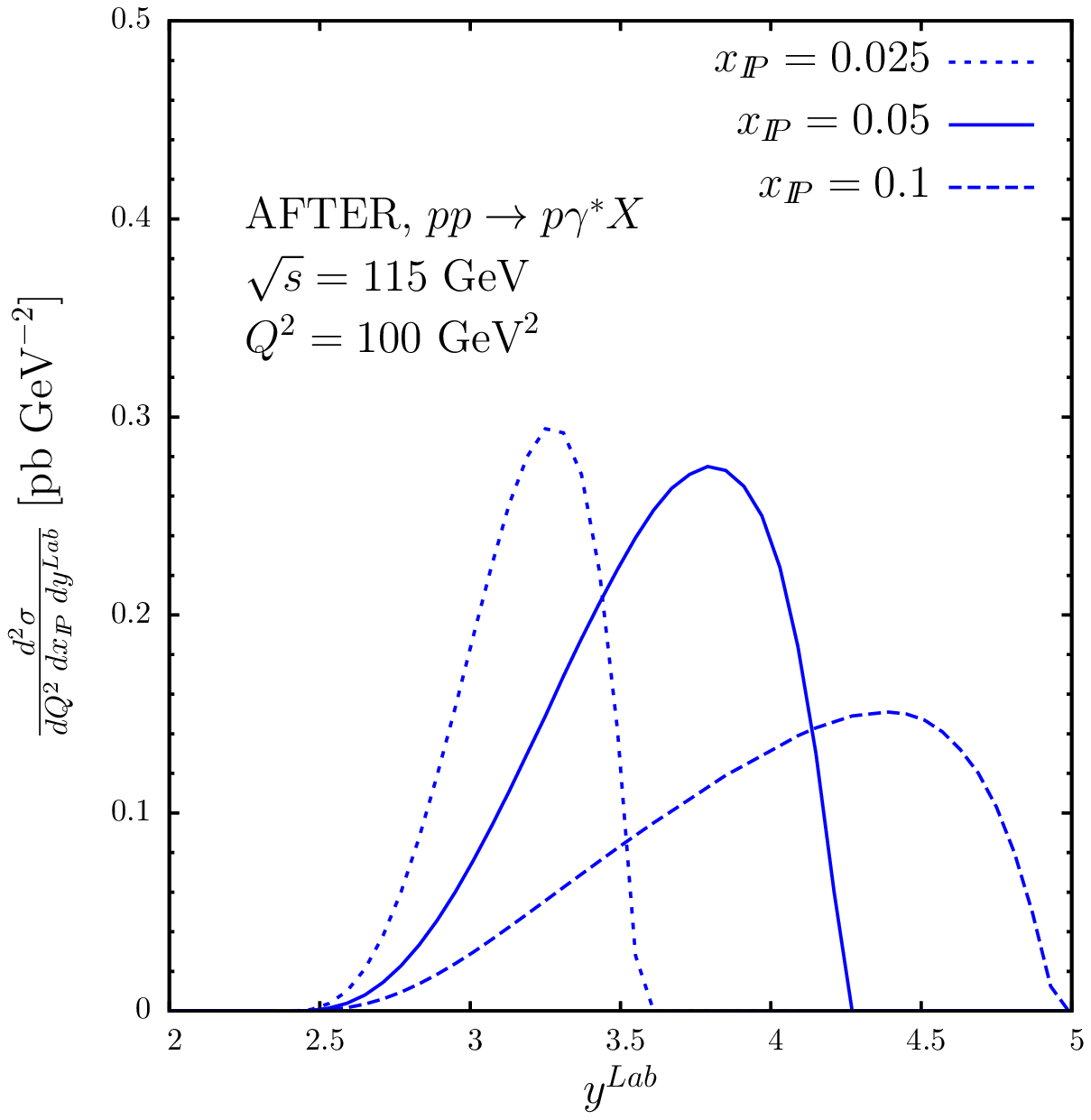}
\caption{\textsl{Left: Triple differential cross sections for the production of a Drell-Yan pair at $x_{\!I\!P}=0.05$
and of mass $Q^2=100$ Ge$\mbox{V}^2$. Blue error bands represent theoretical errors estimation, as described in the text. Right: Triple differential cross sections for three different $x_{\!I\!P}$-values.}}
\label{fig:y}
\end{figure}
The rapidity distributions is particularly sensitive to the shape the diffractive 
parton distributions. This distribution will be useful to investigate any possible kinematic dependence 
of the RGS factor.   
In the left panel of Fig.~(\ref{fig:y}) we present the centre-of-mass rapidity distribution at fixed $Q^2=100$ Ge$\mbox{V}^2$ and $x_{\!I\!P}=0.05$. In this frame the distribution is shifted at negative values of $y^{CM}$.
Therefore, on average, the parton originating from the target proton carries 
more momentum than the one originating from the pomeron.
Since the rapidity is additive under boost along the collision axis we may easily 
boost the $y^{CM}$ to the laboratory frame by using 
\begin{equation}
y^{Lab}=\frac{1+c}{1-c} + y^{CM} \;\;\; \mbox{with} \;\;\;  c=\sqrt{1-\frac{4 m_p^2}{s}}
\end{equation}
with $m_p$ the proton mass. In the AFTER@LHC kinematics this implies 
a rapidity shift $\Delta y = y^{Lab}-y^{CM} = 4.8$.
The rapidity distributions in the laboratory frame for a Drell-Yan pair of mass $Q^2=100$ Ge$\mbox{V}^2$ and 
for three different $x_{\!I\!P}$ values are presented in the right panel of Fig.~(\ref{fig:y}).
One may notice from the plot that for increasing $x_{\!I\!P}$ the Drell-Yan pair spans a wider 
rapidity range and the corresponding spectrum is increasingly more forward. 
It might be useful to discuss the single diffractive Drell-Yan pair production 
in conjuction with the analogous inclusive process. 
\begin{figure}[t]
\includegraphics[scale=0.6]{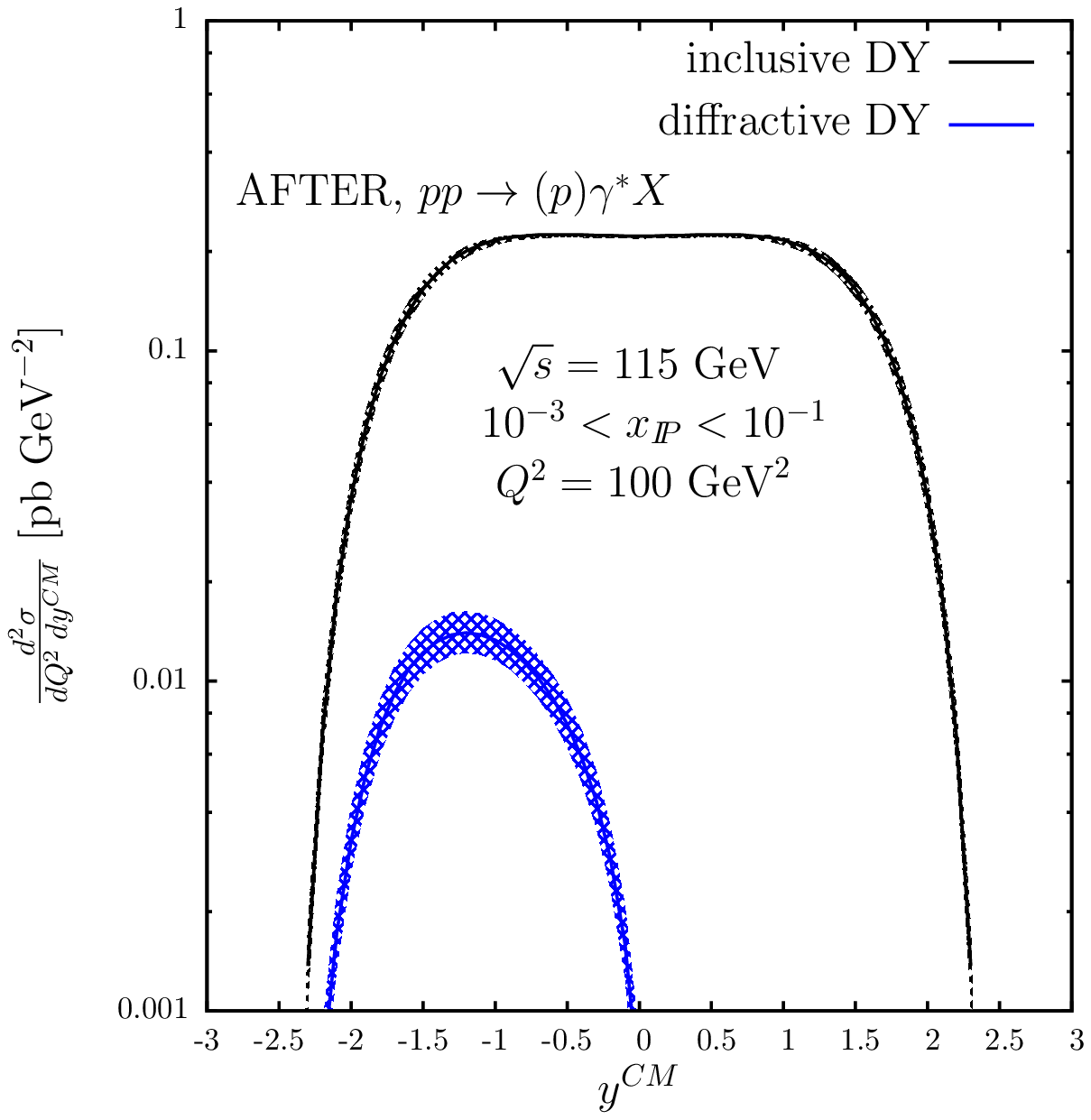}
\includegraphics[scale=0.6]{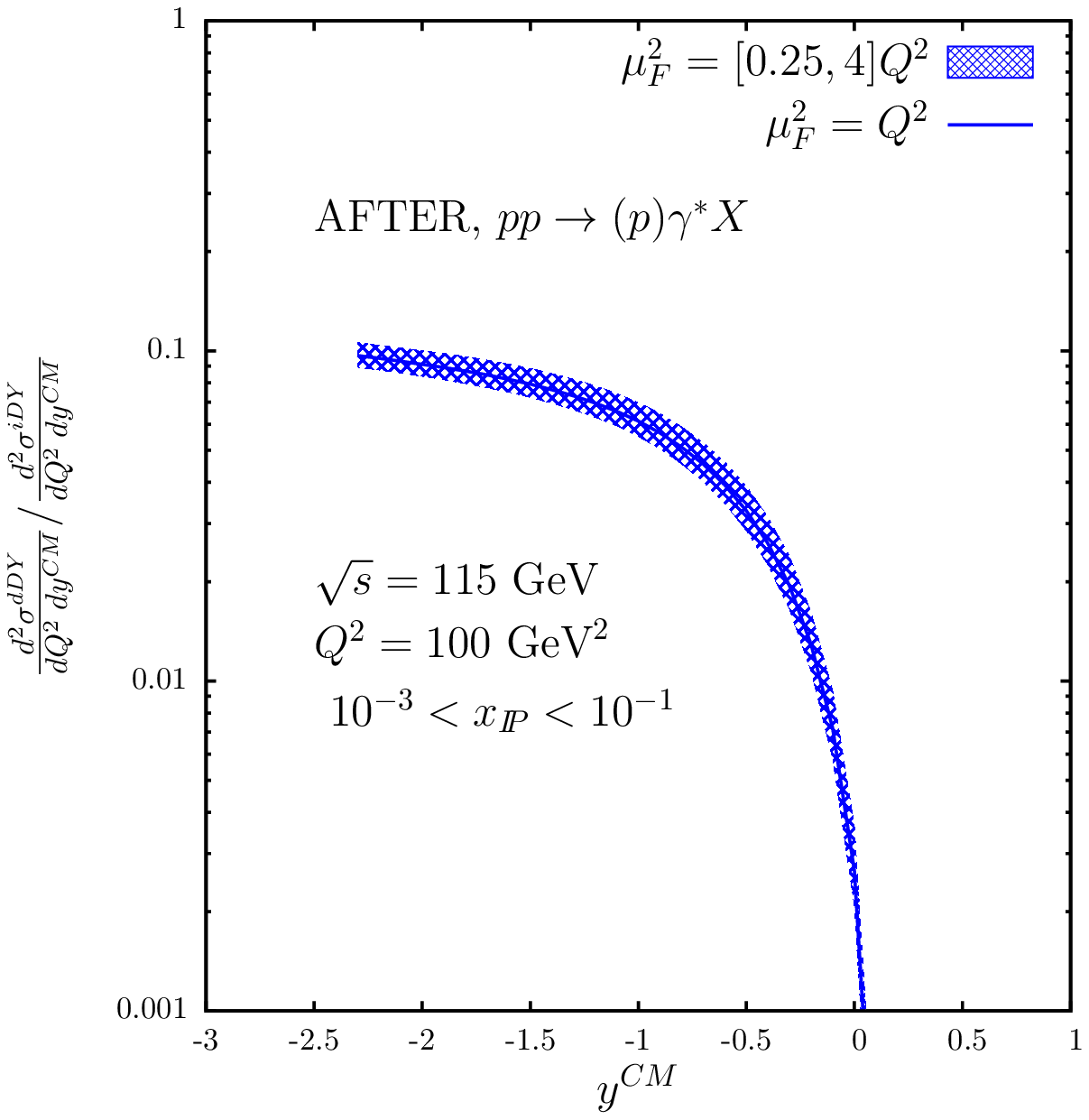}
\caption{\textsl{Left: Rapidity distributions for inclusive and diffractive Drell-Yan of mass $Q^2=100$ Ge$\mbox{V}^2$. Blue error bands represent theoretical errors estimation, as described in the text. Right: diffractive to inclusive  Drell-Yan rapidity distributions ratio.}}
\label{fig:iDY_vs_dDY}
\end{figure}
Such a comparison is presented for the centre-of-mass rapidity distributions in the left plot of Fig.~(\ref{fig:iDY_vs_dDY}) for a common Drell-Yan pair of mass squared $Q^2=100$ Ge$\mbox{V}^2$ and, for the diffractive case, integrated in the range $10^{-3}<x_{\!I\!P}<10^{-1}$. The rapidity distributions in the single diffractive case is strongly asymmetric whereas in the inclusive case it is symmetric around $y^{CM}=0$. This effect is primarily 
due to the different kinematics of the two processes and to the different fractional momentum distributions between 
parton and fracture distributions. 
In the right plot of Fig.~(\ref{fig:iDY_vs_dDY}) the ratio between the two previous distributions is presented. 
The ratio gives direct information on the suppression factor between the single diffractive 
to the inclusive process, assuming a factorised expression for the former, eq.~(\ref{LO_diff_diffy}). 
Such ratio might be convenient from the experimental side since many lepton detection systematics will cancel. On the theoretical one it is expected to be more stable against the inclusion of higher order corrections.  
In the present case, for example, the factorisation scale is simultaneously varied both on the numerator and denominator resulting in a reduced theoretical error band with respect to the one obtained for absolute cross sections.

We wish to end this section we a brief overview of other possible applications of the proposed formalism.
A completely analogous program can be performed for the associated production 
of forward neutron and a Drell-Yan pair,  $p+p\rightarrow n +\gamma^*+X$.
The production of forward neutron in DIS at HERA has shown  
a leading twist nature. From scaling violations of the semi-inclusive 
neutron structure functions a set of proton-to-neutron fracture functions
set has been extracted from data in Ref.~\cite{ceccopieri_neutron}
which can be used to predict forward neutron rate in hadronic collisions.
As in the case of hard diffraction, both physics programs would highly 
benefit from the installation of a dedicated instrumentation for the measuraments 
of fast neutrons and protons quite close to the beam axis. Measuraments in the forward region, althought problematic experimentally, give in fact direct access to the study of the beam fragmentation region.

As a third application we consider hyperon
production associated with a Drell-Yan pair, 
$p+p\rightarrow V+\gamma^*+X$, where $V$ generically indicates 
either a $\Lambda^0$ or $\bar{\Lambda}^0$ hyperon.
At very low transverse momentum, $\Lambda^0$ longitudinal momentum spectrum should show a significant 
leading particle effect, which can be predicted, assuming factorisation as in $d\sigma^{H,T}$, 
by the proton-to-lambda fracture functions set obtained from a fit to SIDIS data in Ref.~\cite{ceccopieri_mancusi}.
On the other hand $\bar{\Lambda}^0$ spectrum in the same kinematical conditions should instead show almost no leading particle effect, giving access to the proton-to-$\bar{\Lambda}^0$ fracture functions.
We note, in general, that the particle-to-antiparticle fracture function is indeed an interesting and almost unknown
distribution. On the other hand, if one considers $\Lambda^0$ or $\bar{\Lambda}^0$ at sufficiently large transverse momentum, their combined analysis, described by $d\sigma^{H,C}$, should 
allow the investigation of parton hadronisation into hyperons in the QCD vacuum as parametrised by fragmentation functions.

As a last application we consider the associated production of one particle and a Drell-Yan pair 
in the context of multi-parton interactions. 
The latter process has already been used to investigate the contamination of the so-called underlying event~\cite{pedestal} to jet observable and has been successfully used to study underlying event properties~\cite{UEDY_CDF}. 
If the detected hadron is measured at sufficiently large
transverse momentum, the latter constitutes a natural infrared regulator for the partonic matrix elements.
In this kinematics conditions we also expect a rather small contributions from fracture functions.  
Therefore the central term,  $d\sigma^{H,C,(1)}$, 
can be used to estimate  the single parton scattering contribution to the process. 
The latter might be considered as the baseline to study the contributions
of double (or multiple) parton scattering contributions to the same final state, 
where, for example, the primary scatter produces a Drell-Yan pair while  
the secondary one produces the detected hadron $H$.  

\section{Conclusions}
\label{Conclusions}
We have briefly reviewed a perturbative approach to single particle production 
associated with a Drell-Yan pair in hadronic collisions. 
On the theoretical side we have shown that the introduction of 
fracture functions allows a consistent factorisation 
of new class of collinear singularities arising in this type of processes.
The factorisation procedure coincides with the one used in 
DIS confirming, as expected, the universal structure of collinear singularities and 
supports the proposed collinear factorisation formula.
On the phenomenological side we have outlined some areas in which the formalism can be fully tested.
In particular, focusing on the AFTER@LHC kinemtic range, 
we have discussed in some detail the single diffractive production of virtual photons.
The study of such a process might improve our understanding of non-perturbative 
aspects of QCD and it allows to explore in detail the nature of factorisation breaking  
at intermediate energies.

\end{document}